# Peculiarities of the electronic transport in half-metallic Co-based Heusler alloys

V.V. Marchenkov[a, b, c*], Yu.A. Perevozchikova[a], N.I. Kourov[a], V.Yu. Irkhin[a], M. Eisterer[d], T. Gao[e]

[a] *M.N. Mikheev Institute of Metal Physics, Ekaterinburg, Russia*
[b] *Ural Federal University, Ekaterinburg, Russia*
[c] *International Laboratory of High Magnetic Fields and Low Temperatures, Wroclaw, Poland*
[d] *Atominstitut, TU Wien, Vienna, Austria*
[e] *Shanghai University of Electric Power, Shanghai, China*

**Abstract**

Electrical, magnetic and galvanomagnetic properties of half-metallic Heusler alloys of $Co_2YZ$ ($Y$ = Ti, V, Cr, Mn, Fe, Ni, and $Z$ = Al, Si, Ga, Ge, In, Sn, Sb) were studied in the temperature range 4.2–900 K and in magnetic fields of up to 100 kOe. It was found that varying $Y$ in $Co_2YZ$ alloys affects strongly the electric resistivity and its temperature dependence $\rho(T)$, while this effect is not observed upon changing $Z$. When $Y$ is varied, extrema (maximum or minimum) are observed in $\rho(T)$ near the Curie temperature $T_C$. At $T \leq T_C$, the $\rho(T)$ behavior can be ascribed to a change in electronic energy spectrum near the Fermi level. The coefficients of normal and anomalous Hall effect were determined. It was shown that the latter coefficient, $R_S$, is related to the residual resistivity $\rho_0$ by a power law $R_S \sim \rho_0^k/M_S$ with $M_S$ the spontaneous magnetization. The exponent $k$ was found to be 1.8 for $Co_2FeZ$ alloys, which is typical for asymmetric scattering mechanisms, and 2.9 for $Co_2YAl$ alloys, which indicates an additional contribution to the anomalous Hall effect. The type of the temperature dependence $\rho(T)$ is analyzed and discussed in the frame of two-magnon scattering theory.

## 1. Introduction

Heusler alloys $X_2YZ$ (where $X$ and $Y$ are transition 3d-elements and $Z$ is an s- or p-element of the Periodic Table) that exhibit half-metallic ferromagnetism are potential candidates for application in spintronics [1, 2]. The main feature of the electronic structure of half-metallic ferromagnets (HMF) is the presence of an energy gap at the Fermi level in one spin sub-band and a metallic character of the density of states in the other [3, 4]. This can lead to 100% spin polarization of the charge carriers, which can be used for spintronic devices. The position and the width of the energy gap can vary quite strongly in different HMF. These parameters can be changed by varying the 3d-, s- and p-elements in $X_2YZ$ Heusler alloys, altering thereby electronic properties.

In 2008, a new class of materials – spin gapless semiconductors – was predicted [5]. These compounds have a number of unique properties associated with their unusual band structure. Such materials enable a combination of the properties of HMF with semiconducting characteristics, a fine regulation of the energy gap, and hence a control of the electrical properties.

The aim of the present work is to study the peculiarities of the electronic transport in Co-based Heusler alloys, namely $Co_2YAl$ and $Co_2FeZ$ with $Y$ = Ti, V, Cr, Mn, Fe, Ni and $Z$ = Al, Si, Ga, Ge, In, Sn, Sb, when the parameters of electronic energy spectrum near $E_F$ can be changed by varying either the 3d- or s- and p-elements. A differ-ence in the electronic transport properties behavior of $Co_2YAl$ and $Co_2FeZ$ systems has been expected.

---

[*] Corresponding author
*E-mail address*: march@imp.uran.ru

## 2. Experimental

The alloys were prepared by arc and induction melting methods in a purified argon atmosphere and annealed at 800 K during 48 h. The atomic content of elements was measured by a scanning electron microscope equipped with an EDAX X-ray microanalysis attachment. Our examination showed that the deviations from a stoichiometric composition were insignificant in all alloys. X-ray diffraction analysis supported the formation of the $L2_1$ structure in the alloys. The structural analysis was performed at the Center of Collective Use, M.N. Mikheev Institute of Metal Physics, UB RAS and Center Nanotech, Ural Federal University.

The magnetic and galvanomagnetic properties were measured at the Atominstitut, TU Wien using an MPMS XL7 (Quantum Design) SQUID magnetometer and a 17 T superconducting solenoid (Oxford). The resistivity and Hall effect were measured by the standard dc four-probe method (see, e.g. Refs [6, 7]) in the temperature range 4.2-900 K and in magnetic fields of up to 100 kOe.

## 3. Results and discussion

Figures 1 and 2 show the dependences of the electrical resistivity of the $Co_2YAl$ alloys (Fig. 1) and $Co_2FeZ$ (Fig. 2), where $Y$ = Ti, V, Cr, Mn, Fe, Ni; and $Z$ = Al, Si, Ga, Ge, In, Sn, Sb. One can see that anomalies are observed for alloys of the $Co_2YAl$ system, when the $Y$ 3d-metal component changes: 1) the value of the residual resistance; 2) the appearance of extrema near the Curie temperature $T_C$; 3) the presence of regions with a negative temperature coefficient of resistance. As shown in Ref. [8, 9], this may be due to the presence of a gap in the electronic energy spectrum near the Fermi level $E_F$.

On the contrary, in the case of $Co_2MeZ$ alloys, when the $Z$-component changes due to the variation of the s- and p-elements, such anomalies are absent. The residual resistivity $\rho_0$ is relatively small, and the temperature dependence $\rho(T)$ has a metallic character. It should be noted that the $T_C$ values of $Co_2FeZ$ alloys are relatively small in comparison with those of $Co_2YAl$ alloys [10-18]. It can be expected that the presence of a gap on $E_F$ should be manifested in other transport properties, in particular in the kinetic coefficients, and, apparently, manifest differently in $Co_2YAl$ and $Co_2FeZ$ alloys. This should be especially noticeable at temperatures much lower than $T_C$. This is why the field dependences of the magnetoresistance and the Hall effect, as well as of the magnetization were studied at liquid helium temperature T = 4.2K.

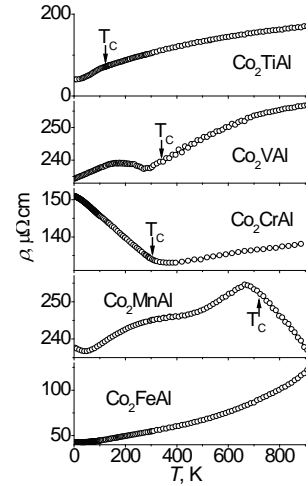

Fig.1. Temperature dependence of electrical resistivity of $Co_2YAl$ alloys. The arrows indicate Curie temperature $T_C$.

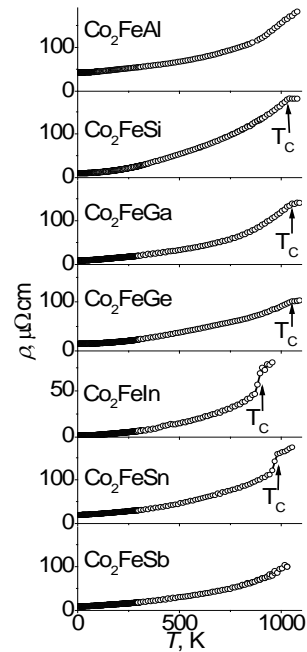

Fig. 2. Temperature dependence of electrical resistivity of $Co_2FeZ$ alloys. The arrows indicate Curie temperature $T_C$.

Since all the alloys are in this ferromagnetic state, their Hall coefficient $R_H$ (the Hall resistance $\rho_{xy}$), besides the normal component $R_0$, contains also the anomalous $R_S$ component. For their separation, we used the method proposed in Ref. [19] when

$$\rho_{xy}(H) = R_0 B + 4\pi R_S M. \tag{1}$$

The first term describes the normal Hall effect. It is due to an action of the Lorentz force on the conduction electrons in a magnetic induction B, which is determined by the external magnetic field $H$ and the magnetization $M$: $B = H + 4\pi M(1 - N)$ and $M = M_S + \chi H$. Here $0 \leq N \leq 1$ is the demagnetization factor of the sample, $\chi$ is the magnetic susceptibility, $M_S$ is the spontaneous magnetization.

From the experiment for both the $Co_2Y Al$ and $Co_2FeZ$ systems, the normal Hall effect (NHE) coefficient $R_0$ the anomalous Hall effect (AHE) coefficient $R_S$, the residual resistivity $\rho_0$, and the spontaneous magnetization $M_S$ were determined (Figures 3 and 4).

As shown in Fig. 3, a large change in the coefficient $R_0$ is observed in $Co_2Y Al$ depending on the number of valence electrons per formula unit $z$. First of all, the sign of the coefficient $R_0$ changes from positive to negative by the transition from alloys with $z < 27$ to alloys with $z \geq 27$ with a sufficiently large value of the spontaneous magnetization $M_S$. This may indicate a change in the carrier type at $z \sim 27$. In addition, the absolute value of the coefficient $R_0$ increases significantly in alloys with the $Y$ component from the middle of the 3d period. These features of the $R_0(z)$ behavior indicate a corresponding rearrangement of the electron spectrum near $E_F$ with a change in $z$.

On the other hand, Fig. 3 shows that in the middle of the 3d period the residual electrical resistivity $\rho_0$ also increases sharply (at T = 4.2 K). It seems that the observed correlation in the behavior of the dependences of $R_0(z)$ and $\rho_0(z)$ indicates the essential role of scattering processes of current carriers in the formation of AHE. However, as was shown earlier [6-8], the electrical resistivity of the alloys considered is mainly determined by the parameters of the electron spectrum at $E_F$. We can thus conclude that the changes in $R_0$ and $\rho_0$ depend on the number of valence electrons $z$ due to the same reasons, namely, a rearrangement of the electronic band structure near $E_F$.

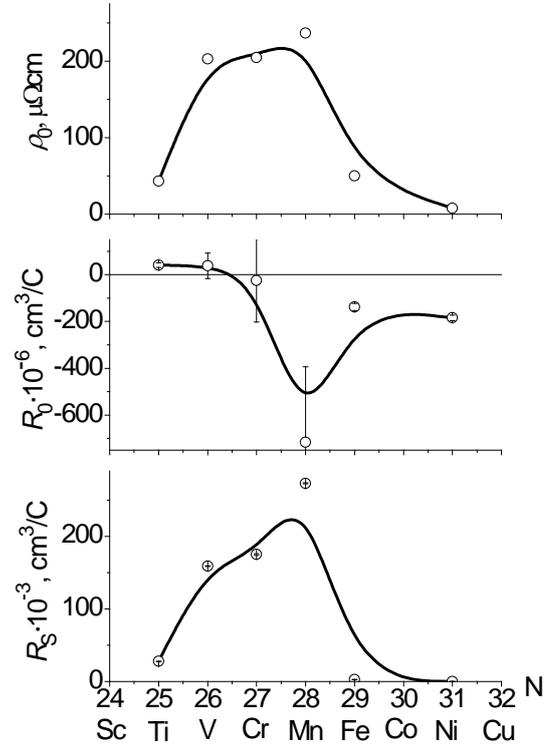

Fig. 3. Dependences of $\rho_0$, $R_0$, $R_S$ in $Co_2Y Al$ alloys on the number of valence electrons.

The coefficient of $R_S$ of ferromagnetic alloys is related to their resistivity $\rho$ and the spontaneous magnetization $M_S$ by [20, 21]

$$R_S \approx \lambda_{\text{eff}} \cdot \frac{\rho^k}{M_S}, \tag{2}$$

where $\lambda_{\text{eff}}$ is the effective spin-orbit interaction (SOI) parameter, and $k$ is an exponent whose value depends on the scattering mechanism of the charge carriers. Usually, the value of $k$ is 1 or 2.

From a comparison of the experimental data presented in Fig. 3, one can see that in the investigated alloys the value of the coefficient $R_S$ is almost three orders of magnitude higher than that of $R_0$. The AHE coefficient has a positive sign for all the alloys. In weakly ferromagnetic alloys with $z < 27$, the sign of the coefficient $R_S$ is opposite to the sign of $R_0$. According to the relation (2), the observed difference in the sign of the AHE and NHE can be related for any scattering mechanism to the sign of $\lambda_{\text{eff}}$. It is noteworthy that the AHE coefficient increases sharply in the middle of the 3rd period, as well as the values of $R_0$, $\rho_0$ и $M_S$.

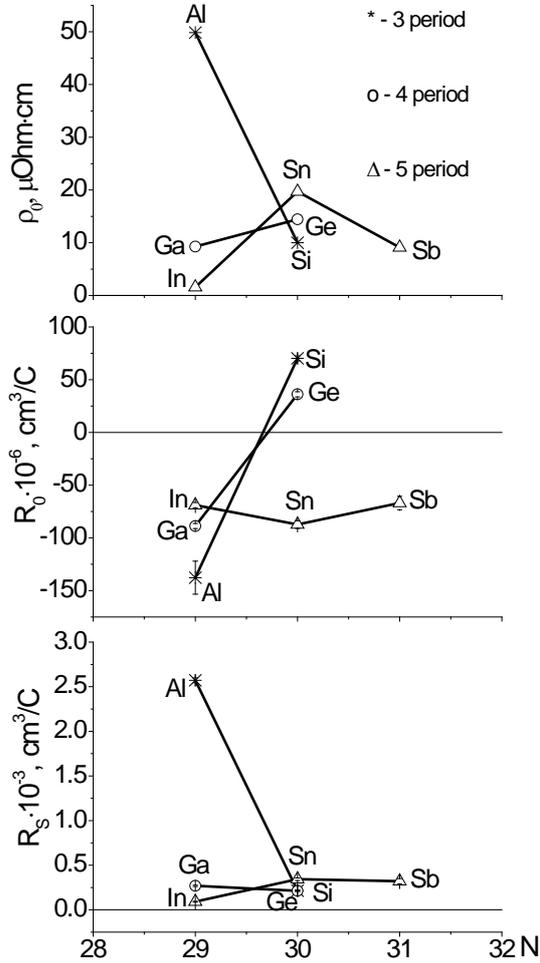

Fig. 4. Dependence of $\rho_0$, $R_0$ and $R_S$ in Co$_2$FeZ alloys on the number of valence electrons.

One of the most important questions in the analysis of the AHE is the determination of the main mechanism of carrier scattering. It is known [20, 21] that the dependence of $R_S$ on $\rho_0$ can be quadratic in the case of: 1) the proper SOI (interaction of the spin of the electron with its own orbital motion); 2) an improper SOI (interaction of the electron spin with the orbital motion of another electron); 3) the mechanism of side-jump scattering, i.e., a discontinuous change in the trajectory of an electron scattered by an impurity. In addition, the mechanism of asymmetric scattering is usually considered in the analysis of the AHE, when the probability of scattering of an electron to the left or to the right from its motion direction is assumed to depend on the spin direction for a proper or improper SOI [22].

As one can see from Fig. 5, the $R_S$ coefficient determined at T = 4.2 K has a power dependence (2) in the Co$_2$YAl system with the exponent $k$ = 2.9. Apparently, none of these known models, commonly used in the analysis of the AHE, does adequately describe it in these alloys. It should be noted that the dependence of type (2) with the exponent $k$ = 3.1 was obtained for a similar Fe$_2$YAl system in [19]. Most likely, the AHE of the alloys under consideration is to a large extent determined not so much by the scattering mechanisms of the current carriers in the presence of a periodic SOI, as by the rearrangement of the electronic band structure near the Fermi level $E_F$, accompanied by a change in the number of current carriers. Therefore, it seems interesting to study such dependences in the Co$_2$FeZ system, where the states of valence electron number changes due to Z elements which are s- and p-elements of the 3rd, 4th and 5th periods of the Mendeleev table.

Fig. 4 shows the dependences of $\rho_0$, $R_0$ and $R_S$ on the number of valence electrons in Co$_2$FeZ alloys. It can be seen that $R_S$ for all the alloys is much higher than $R_0$. In addition, for most alloys the sign of the coefficient $R_S$ is opposite to the sign of the coefficient $R_0$, which, apparently, is due to the negative sign of $\lambda_{eff}$.

To verify the validity of relation (2), the dependences $R_S = f(\rho_0)$ were investigated. As seen in Fig. 6, a power-law dependence of $R_S$ on $\rho_0$ with the exponent $k$ = 1.8 is observed in all the investigated alloys. Hence, for the half-metallic Co$_2$FeZ system, where the Z elements vary, an additional contribution to the AHE caused by the change in the number of current carriers is practically absent because of a slight change in the parameters of the electron spectrum near $E_F$. In this system, the change in the magnitude of both $R_S$ and $\rho_0$ as a function of the Z component is due to the scattering mechanisms only. A comparison of the results of the AHE study in the Co$_2$FeZ HMF system with the experimental data for the Fe$_2$YAl [19] and Co$_2$YAl systems (Figure 5) confirms the point of view suggested in [20, 21] that one needs to take into account the additional contribution to the AHE caused by a change in the number of the current carriers.

Among the HMF Heusler alloys studied, the Co$_2$FeSi alloy with large $T_C$ = 1100 K is of particular interest. According to [23], a $T$-exponential contribution to $\rho(T)$ can be extracted in single crystals of this compound. In the opinion of the authors of [23], this is a manifestation of the energy gap for current carriers with spin down. However, the values of ρ for the investigated crystal are small, typical for metals, and $\rho(T)$ has an "ordinary" metallic form, whereas the energy gap is rather large. According to [24], the behavior of magnetic resistivity is determined by two-magnon scattering processes in such a situation. They lead to a characteristic dependence $\rho(T) \sim T^n$ with $7/2 < n < 9/2$ depending on temperature and negative linear magnetoresistance $\Delta\rho_{xx} \sim H^{-1}$.

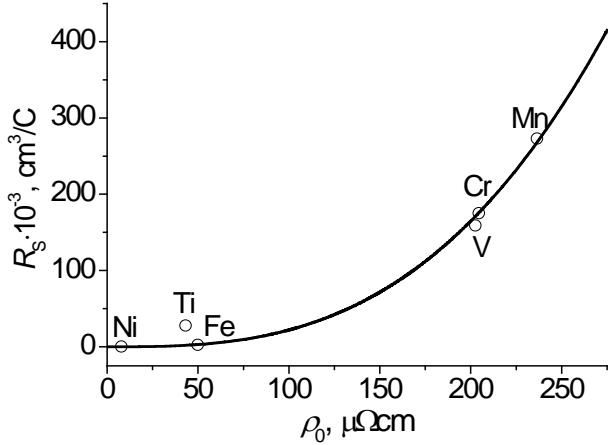

Fig. 5. Dependence of the $R_S$ coefficient on the residual resistivity $\rho_0$ of $Co_2YAl$ alloys at T = 4.2 K.

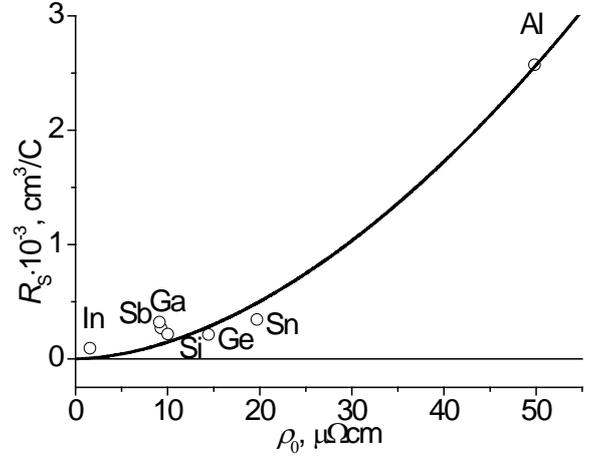

Fig. 6. Dependence of the $R_S$ coefficient on the residual resistivity $\rho_0$ of $Co_2FeZ$ alloys at T = 4.2 K.

As a result of our study, it was found that there are three temperature intervals where the resistivity depends on temperature and magnetic field in different ways: (a) below 30 K $\rho(T) \sim T^n$, where $n \approx 2$, and the magnetoresistance is $\Delta\rho_{xx} > 0$; (b) from 30 to 80 K $\rho(T) \sim T^n$, where $n \approx 7/2$, and $\Delta\rho_{xx} \sim H^{-1}$; (c) above 80 K $\rho(T) \sim T^n$, where $n \approx 2$, and $\Delta\rho_{xx} < 0$. The experimental results obtained indicate that a power-law temperature dependence of the electrical resistivity with the exponent $n \approx 7/2$ and a linear negative magnetoresistance prevails in the temperature range $30 K < T < 80 K$. This appears to be a manifestation of two-magnon scattering processes as the main mechanism for the scattering of current carriers, which determines the behavior of the resistivity and magnetoresistivity of the $Co_2FeSi$ alloy at temperatures $30 K < T < 80 K$. At the same time, at higher temperatures one-magnon processes have to be considered.

## Conclusions

We have demonstrated that the variation of $Y$ strongly affects the number of current carriers and alters the electronic band structure near the Fermi level $E_F$, and, hence, the electronic properties of $Co_2YAl$. In case of $Co_2FeZ$ alloys, varying $Z$ does not greatly change the electronic transport properties. Analysis of obtained results allows us to suppose that the $Co_2FeZ$ system, in particular, the $Co_2FeSi$ alloy, is more preferable for using in spintronics than the $Co_2YAl$.

## Acknowledgements

This work was partly supported by the state assignment of FASO of Russia (theme "Spin" No. 01201463330), by the Scientific Program of UB RAS (project No. 15-17-2-12), by the RFBR (grant Nos. 15-02-06686), by the Government of the Russian Federation (state contract No. 02.A03.21.0006), by the National Natural Science Foundation of China (No. 11204171) and by the Shanghai Municipal Natural Science Foundation (No. 16ZR1413600).